# Photonic chiral bulk transports manipulated by boundary freedom in three-dimensional meta-crystals

*Yingxin Qi[1,2], Hanyu Wang[1,2], and Qinghua Guo[3]\*, Zhihong Zhu[1,2]\*, Biao Yang[1,2]\**

[1] College of Advanced Interdisciplinary Studies &Hunan Provincial Key Laboratory of Novel Nano-Optoelectronic Information Materials and Devices, National University of Defense Technology, Changsha 410073, China

[2] Nanhu Laser Laboratory, National University of Defense Technology, Changsha 410073, China.

[3] School of Physics and Electronics, Hunan University, Changsha 410082, China

*Corresponding to: B. Yang (yangbiaocam@nudt.edu.cn); Z. Zhu (zzhwcx@163.com);

Q. Guo(guoqh@hnu.edu.cn)

**Keywords:** boundary-bulk correspondence, chiral bulk transports, zeroth modes, nodal line meta-crystals

**Abstract**

In topological physics, one of the most intriguing phenomena is the presence of topological boundary states, accurately predicted by the well-established bulk-edge correspondence. For example, in three-dimensional Weyl semimetals, Fermi arcs emerge to connect projected Weyl points on the surface due to inheriting the bulk-edge correspondence from the integer quantum Hall effect. However, limited attention has been paid to exploring the reverse mechanism in topological crystals. In this study, we propose that boundaries can serve as an alternative degree of freedom to manipulate topological bulk transports. We analytically and experimentally validate our concept using a finite-thickness photonic meta-crystal that supports bulk nodal lines, with its zeroth modes exhibiting opposite chiral bulk transports under different boundary conditions. Notably, the mirror symmetry remains preserved across both configurations. These findings are applicable to other topological systems, providing new insights into systems with varied boundary conditions and offering the potential for the design of more compact and spatially efficient topological photonic devices.

**Introduction**

Topological semimetals represent a distinct class of topological states, separate from topological insulators, and have emerged as a prominent research focus in recent decades. Based on various configurations of energy band crossings in momentum space, topological semimetals can be categorized into types such as Dirac semimetals, Weyl semimetals, nodal line semimetals, and other unconventional semimetals[1–10]. Among these, Weyl semimetals are particularly representative within the field of topological physics[11–20]. In bulk materials, Weyl points act as monopoles of Berry curvature, while on the surface, Fermi arcs connect the projected positions of these Weyl points, governed by the bulk-edge correspondence, where the topological properties of the bulk dictate the existence of boundary states—a foundational principle in topological physics[21–28].

In other words, the number of surface state arcs directly corresponds to a topological bulk integer known as the Chern number, uniquely defined by the bulk states under periodic boundary conditions. This relationship is robust across various surfaces, where surface states decay exponentially into the bulk. Typically, bulk materials are either considered to be half-infinite or sufficiently thick slabs to ensure complete decay of surface states into the bulk[29–32]. However, when the slab thickness is finite, coupling between the decaying surface waves on opposite surfaces can occur, ultimately generating topological chiral bulk states, which has recently attracted considerable attention[33–36]. The topological chiral bulk states arise from the chiral zeroth Landau level, which is driven by the quantization of energy resulting from the interaction between the three-dimensional Weyl degeneracy and an axial magnetic field or an artificial gauge field[37–39]. Recently, studies have shown that chiral bulk states can be induced by imposing specific boundary conditions on a two-dimensional (2D) Dirac semimetal[40–44]. Despite these insights, the underlying mechanisms related to the inherent topological features arising from different boundary conditions remain underexplored. For instance, whether there exists a degree of freedom associated with boundary conditions that could reveal these topological features remains an open question.

Here, for brevity, we begin with photonic nodal line semimetals and introduce the mechanism for boundary manipulation of topological chiral bulk states. The versatility and adaptability of sample design and fabrication make classical waves an ideal platform for exploring topological physics [45–47]. We analytically and experimentally demonstrate that a truncated photonic nodal line meta-crystal, with boundaries on the top and bottom surfaces, reveals distinct classes of topological chiral bulk states induced solely by boundary conditions, independent of slab thickness. The two chiral bulk states originate from the nodal ring. In the absence of boundaries, the chiral bulk states coexist near the nodal ring and propagate in opposite directions; however, after truncation, these pseudospin-locked states propagate exclusively in a chiral manner—either inwards or outwards, depending on the imposed boundary conditions. Thus, these states are referred to as chiral bulk states. In the analytical solution, we find that the topological chiral bulk states are two independent zeroth-order solutions, completely different from trivial higher-order bulk states. Unlike traditional bulk-edge correspondences, where boundary states are determined by bulk properties, these chiral bulk states challenge conventional wisdom by being governed by boundary conditions. In other words, by manipulating the boundary conditions, we can observe different chiral bulk states. Specifically, positive and negative dispersion chiral bulk states are induced by perfect electric conductor (PEC) and perfect magnetic conductor (PMC) boundary conditions, respectively.

**Results**

The photonic nodal ring in the three-dimensional (3D) momentum space is shown in Figure 1a, formed by the crossing between a mirror-odd mode (red), with a normal electric field $E_z$, and a mirror-even mode (blue), with tangential electric field $E_{x-y}$. It is topologically protected, and its topological invariant is the winding number $w = \pm 1$, which is defined along the red ring around the nodal ring, as shown in the inset of Figure 1a. The configuration of a nodal line meta-crystal with finite thickness can be considered as a photonic meta-crystal slab waveguide, where the upper and lower boundaries act as hard surfaces perpendicular to the z-axis. Due to the finite size effect, the nodal ring is gapped because of the discretized $k_z$, creating a pseudo-gap.

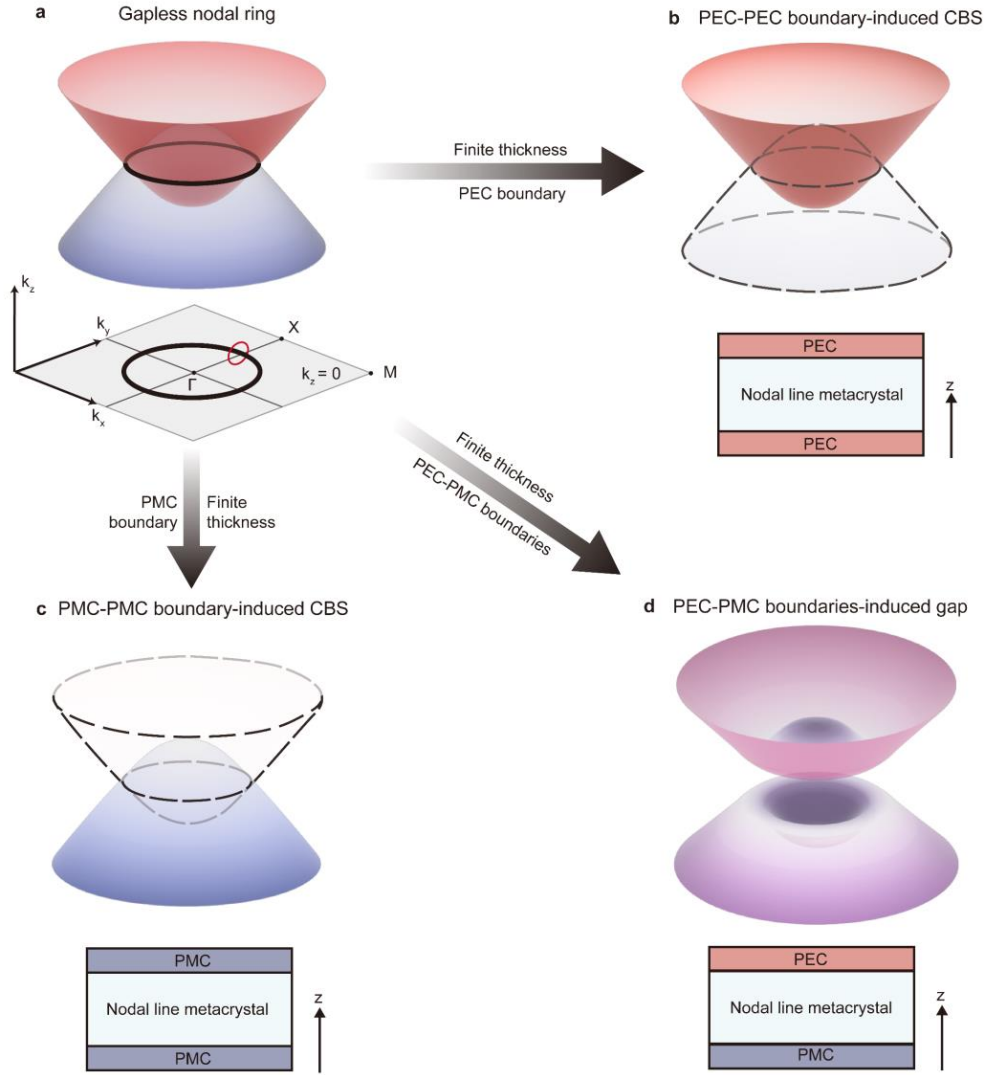

**Figure 1. Schematic diagram of boundary-induced chiral bulk states in a nodal line semimetal. (a)** Gapless nodal ring protected by mirror symmetry $M_z$ in the 3D momentum space. The inset shows the nodal ring (black) in the plane $k_z = 0$ of the Brillouin zone, protected by the winding number $\pm 1$ of the encircling loop (red). **(b, c)** PEC/PMC boundary-induced chiral bulk states in the meta-crystal with finite thickness, where dashed lines indicate forbidden modes. **(d)** Gap induced by the PEC-PMC boundary condition in the meta-crystal. The purple modes result from a hybridization of mirror-odd (red) and mirror-even (blue) modes. The insets in **(b-d)** are the geometric diagrams of the meta-crystal with finite thickness under different boundary conditions.

Figure 1b shows that when both the upper and lower surfaces are PEC boundaries, the mirror-even mode (dashed lines) is forbidden, while the mirror-odd mode remains intact, as PEC boundaries permit only the normal electric field component $E_z$ at the boundaries. When the boundaries are changed to PMC, as shown in Figure 1c, the situation is reversed: PMC boundaries only support the tangential field component $E_{x-y}$, allowing only the mirror-even mode to persist, while the mirror-odd mode (dashed lines) disappears.

Thus, when constructing a finite-size waveguide with PEC or PMC boundaries on both sides, chiral bulk states emerge within the pseudo-gap of the projected bands, as schematically shown in Figures 1b and 1c. Notably, the PEC-PEC and PMC-PMC boundary conditions preserve the system's symmetry, and their differing electromagnetic responses determine the "chirality" of the topological chiral bulk states, which exhibit positive and negative dispersion, respectively. Furthermore, these chiral bulk states are bulk states induced solely by boundary conditions, extend throughout the entire waveguide, and remain unaffected by the waveguide thickness. However, when PEC and PMC boundaries are imposed on the top and bottom surfaces, respectively, the mirror symmetry of the nodal ring is broken, and the chiral bulk states disappear, leading to the formation of a complete energy gap, as illustrated in Figure 1d. In other words, the chiral bulk states arise from the interplay between the nodal ring's topological structure and the presence of mirror-symmetric boundaries. These phenomena across the three different boundary conditions clearly demonstrate that boundary conditions can control the topological transports of bulk states.

*Construction of boundary-induced chiral bulk states*

As a platform for experimental implementation, we selected a nodal line meta-crystal composed of I-shaped metallic resonators, as shown in Figure 2a[48]. To insert the source and probe antennas, we drilled periodic holes into the structure. The unit cell in Figure 2a has a dimension of $10 \times 10 \times 4 \text{ mm}^3$. Two metallic resonators (copper, red) are placed orthogonally on the x-y plane and embedded in a dielectric background with permittivity $\varepsilon = 4.2$ (FR4 materials). The structural parameters are as follows: $a = 10 \text{ mm}, h = 4 \text{ mm}, r_0 = 1.8 \text{ mm}, r = 2.1 \text{ mm}, d = 0.2 \text{ mm}, l = 8.6 \text{ mm}, l_0 = 6.4 \text{ mm}, l_1 = 1.3 \text{ mm}$ and $w = 0.5 \text{ mm}$.

The corresponding bulk band structure is shown in Figure 2b. On the $k_z = 0$ plane, the nodal ring around 4.6 GHz is formed by the linear intersection of the transverse mode (TM) and the longitudinal mode (LM). In terms of mirror symmetry $M_z$, TM (LM) exhibits odd (even) mirror eigenvalues as shown in Figure 2c. The negative dispersion of LM is achieved through an on-local effect by introducing in-plane glide symmetry (i.e. p4g), where the electric field lies within the x-y plane. The TM in the bottom panel of Figure 2c exclusively exhibits an $E_z$ field component, while the $E_x$ and $E_y$ components are nearly negligible.

To simulate the chiral bulk states, 5 layers of unit cells are stacked in the z-direction as shown in Figure 2d, where the PEC boundary is applied on both the top and bottom surfaces to form a meta-crystal waveguide. In this configuration, only the TM mode survives. Conversely, applying a PMC boundary selects the LM mode instead. In particular, for both cases, we only imposed the boundaries, which did not change the mirror symmetry of the system. As a result, the boundaries induce chiral modes (red or blue) to appear in the waveguide, passing through the pseudo-gap (shaded region in Figures 2f and 2g). As shown in Figures 2h and 2i, these chiral modes fully inherit the properties of the TM and LM modes at the nodal ring, respectively.

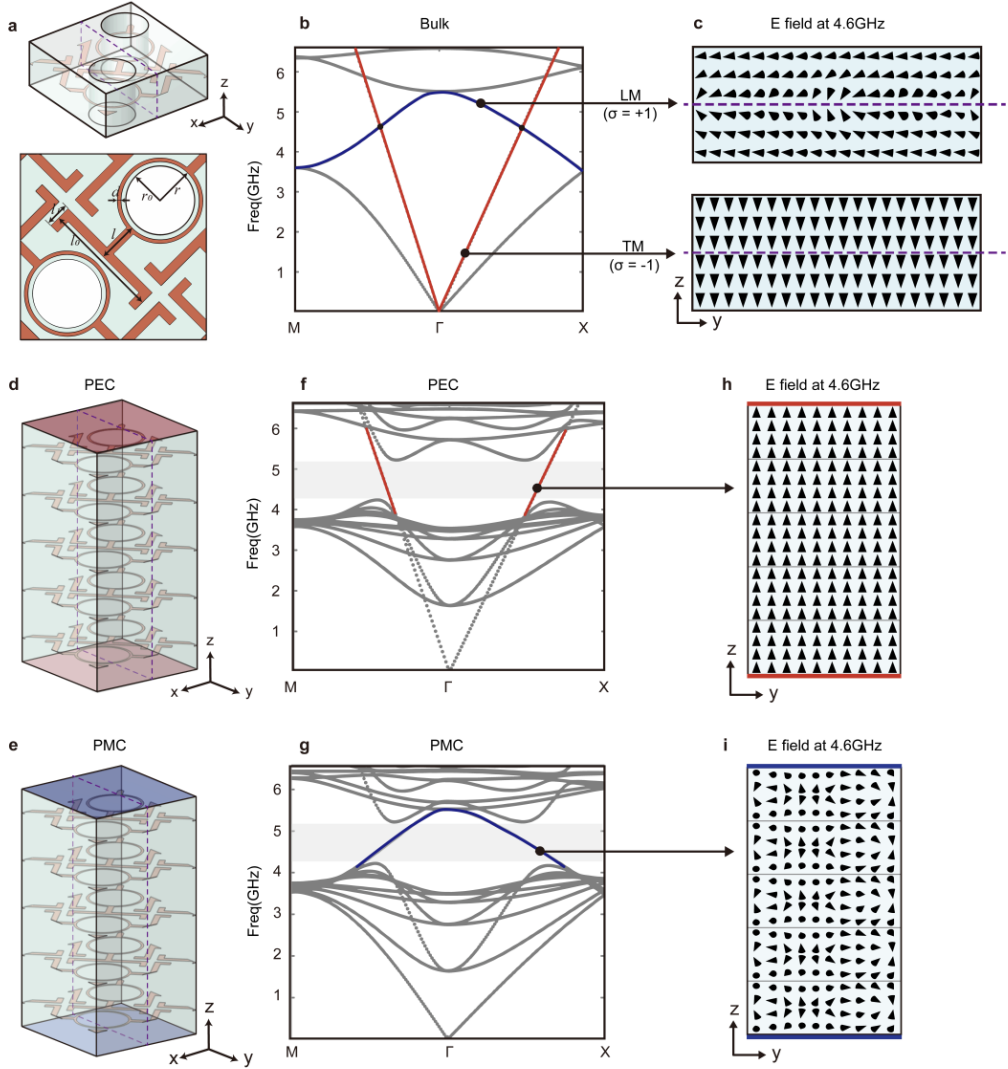

**Figure 2. Boundary-induced chiral bulk states in the nodal line waveguide composed of 5 layers of unit cells. (a)** Geometry and parameters of the unit cell. **(b)** Bulk band diagram. **(c)** Electric field distributions at 4.6GHz on **(a)** the y-z cutting plane (dashed line). **(d, e)** Supercells of waveguides. Red and blue represent PEC and PMC boundaries, respectively. **(f, g)** Band structures of **(d, e)** the supercells. **(h, i)** Electric field distributions at 4.6 GHz of **(f, g)** chiral bulk states (red or blue) on **(d, e)** the y-z cutting plane (dashed line), respectively.

When the number of waveguide layers increases, the pseudo-gap becomes narrower, as shown in Supporting Information Note 1. However, regardless of thickness, as long as the boundary conditions remain unchanged, the chiral bulk states continue to propagate throughout the entire waveguide. Thus, they are largely independent of thickness, which is different from the trivial bulk states (gray). From this perspective, we can imagine that as the waveguide approaches infinite thickness, the pseudo-gap shrinks to an infinitely small size. The chiral bulk states would still exist at this limit, as their existence is solely determined by the boundary conditions, making this a smooth transition.

It is well established that finely tuning the boundary potential in quasi-one-dimensional monolayer graphene can produce chiral states[40]. However, in photonics, hard boundaries such as PEC and PMC exhibit opposite electromagnetic responses[49], which can directly and selectively induce the emergence of different chiral bulk states.

*Analytical solutions*

To investigate the effect of boundaries on the nodal ring, we present analytical solutions of its dispersion characteristics. The effective media of nodal ring can be accurately described by the following constitutive matrices as ($\mathbf{D} = \overleftrightarrow{\epsilon}\mathbf{E}$ and $\mathbf{B} = \overleftrightarrow{\mu}\mathbf{H}$),

$$\overleftrightarrow{\epsilon} = \begin{pmatrix} \epsilon & 0 & 0 \\ 0 & \epsilon & 0 \\ 0 & 0 & 1 \end{pmatrix}, \overleftrightarrow{\mu} = \begin{pmatrix} 1 & 0 & 0 \\ 0 & 1 & 0 \\ 0 & 0 & 1 \end{pmatrix} \tag{1}$$

with,

$$\epsilon = 1 + \frac{l^2}{L(\omega_0^2 - \omega^2)}, l = 1 - \frac{a(k_x^2 + k_y^2)}{b + c(k_x^2 + k_y^2)} \tag{2}$$

where $l$, $L$, $\omega_0$ are effective length, inductance and resonance frequency, respectively. The derivation detail is provided in Supporting Information Note 2. For simplicity, we assume $k_x = 0$. For PEC boundary, the eigenmode dispersions are,

$$
\begin{cases}
\epsilon \neq 0 & (3a) \\
\omega = \pm k_y & (3b) \\
2d\sqrt{\epsilon}\sqrt{\omega^2 - k_y^2} = 2n\pi, n \in \mathbb{Z} \text{ and } n \neq 0 & (3c) \\
2d\sqrt{\omega^2 \epsilon - k_y^2} = 2n\pi, n \in \mathbb{Z} \text{ and } n \neq 0 & (3d)
\end{cases}
$$

For PMC boundary, the eigenmode dispersions are,

$$
\begin{cases}
\epsilon = 0 & (4a) \\
\omega \neq \pm k_y & (4b) \\
2d\sqrt{\epsilon}\sqrt{\omega^2 - k_y^2} = 2n\pi, n \in \mathbb{Z} \text{ and } n \neq 0 & (4c) \\
2d\sqrt{\omega^2 \epsilon - k_y^2} = 2n\pi, n \in \mathbb{Z} & (4d)
\end{cases}
$$

When we consider top-PEC and bottom-PMC boundaries, we have,

$$
\begin{cases}
2d\sqrt{\epsilon}\sqrt{\omega^2 - k_y^2} = (2n+1)\pi, n \in \mathbb{Z} & (5a) \\
2d\sqrt{\omega^2 \epsilon - k_y^2} = (2n+1)\pi, n \in \mathbb{Z} & (5b)
\end{cases}
$$

We observe that for the PEC-PEC and PMC-PMC boundary conditions, the round-trip phase conditions of guide modes are $2n\pi$, while for the PEC-PMC boundary condition, they change to be $(2n+1)\pi$. It is due to that the PEC boundary provides an extra $\pi$ phase delay in the PEC-PMC boundary configuration. It is worth highlighting that for the odd integer condition, i.e. $(2n+1)\pi$, there is no unique order such as the chiral zeroth order.

If we set $a = 1, b = 10, c = 1, L = 1, \omega_0 = 2$, the projected bands are shown in Figure 3. For the equation $2d\sqrt{\epsilon}\sqrt{\omega^2 - k_y^2} = 2n\pi$ ($n \in \mathbb{Z}$), when $n = 0$, there are two independent solutions corresponding to the chiral zeroth modes, i.e. $\epsilon = 0$ and $\omega = \pm k_y$, which cross to form the nodal ring (or Dirac point on the $k_x = 0$ plane). In contrast, the solutions for $n \neq 0$ correspond to the gapped trivial bulk bands. The PEC or PMC boundary condition only allows one to be valid (see Equations 3 and 4) as shown in Figure 3a-f, aligning with the above simulation results. Figures 3a-c (or Figures 3d-f) demonstrate that as the thickness d increases the gap gradually becomes shrinker, which matches the expected behavior.

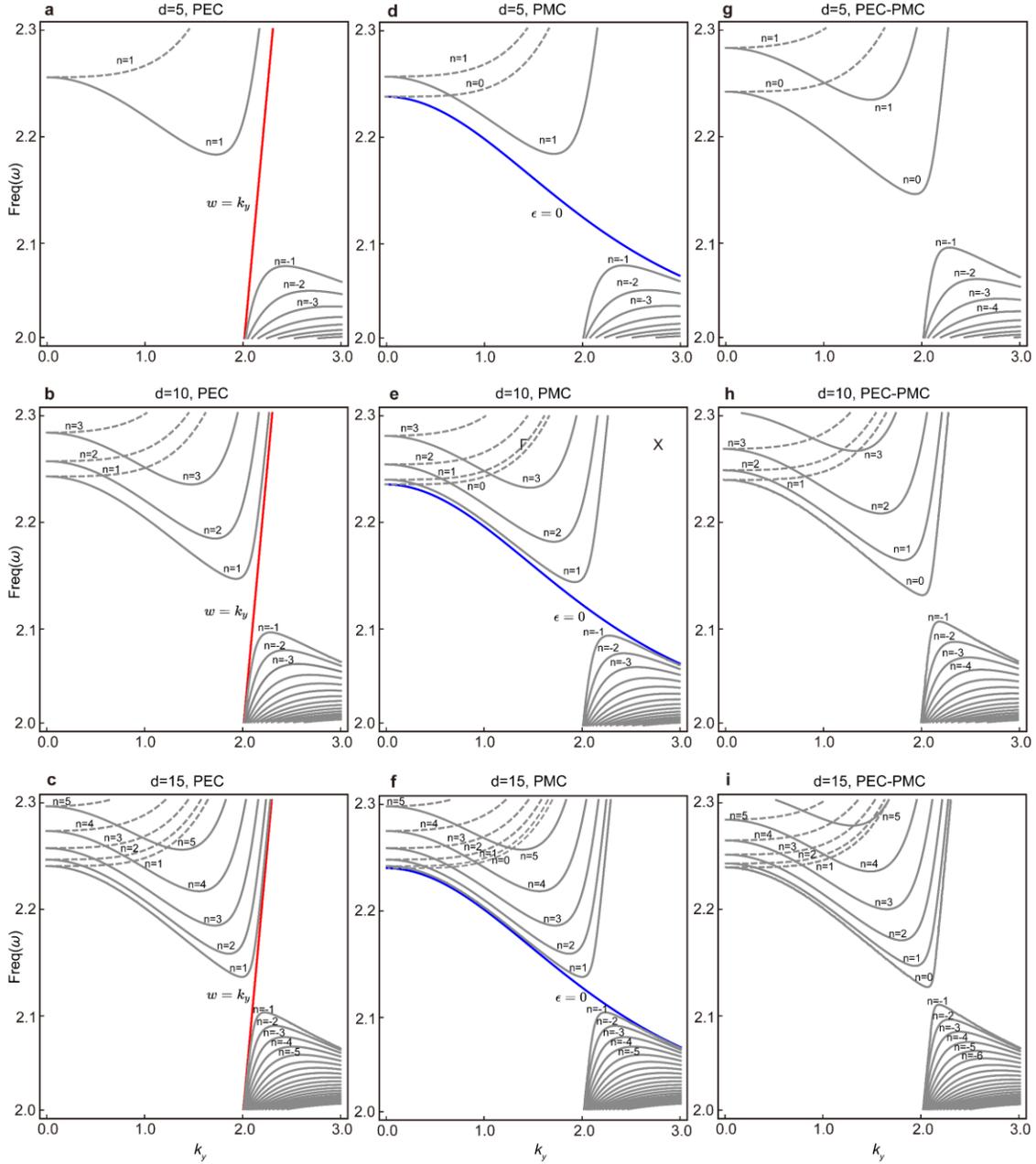

**Figure 3. Analytical solutions of chiral bulk bands. (a-f)** The projected bands along $k_y$. Chiral bulk states (red and blue) under the **(a-c)** PEC-PEC and **(d-f)** PMC-PMC boundary conditions, respectively. Gray solid lines are solutions to Equations 3c and 4c, while gray dashed lines correspond to solutions to Equations 3d and 4d. **(g-i)** PEC-PMC boundary condition. Gray solid lines are solutions to Equation 5a and gray dashed lines are solutions to Equation 5b.

The solutions of $n = 0$ do not explicitly involve $d$, indicating that the topological chiral bulk states remain intact as $d$ changes. The solutions for $n \neq 0$ are $d$-dependent, resulting in a shrinking pseudo-gap as $d$ increase. The bulk states under the PEC boundary (Figures 3a-c) and PMC boundary conditions (Figures 3d-f) are roughly similar, as also described by Equations 3 and 4. However, they are significantly different from those under the top-PEC and bottom-PMC boundary condition as shown in Figures 3g-i, which is due to the boundary effects.

In essence, the nodal ring ensures the theoretical existence of chiral bulk states, as confirmed by our analytical solutions. As shown in Supporting Information Note 3, when the waveguide thickness $d$ is significantly large, the trivial high-order modes remain gapped, while the chiral zeroth-order modes persist, finally approaching the bulk nodal lines (see red and blue lines).

*Experimental observation of chiral bulk transports*

Subsequently, we experimentally demonstrated the chiral bulk transport properties in the nodal line waveguides. Figure 4 shows the artificial magnetic conductor (AMC), which serves as an experimental alternative to both the PMC and PEC. Further details are provided in Supporting Information Notes 4 and 5. The AMC is an artificial periodic structure composed of a ground plane, dielectric substrate, metal through-holes, and upper metal patches (see Figure 4a). As shown in Figure 4b, when the patch is positioned close to the meta-crystal, it exhibits reflective properties for electromagnetic waves within a specific frequency range, with phase behavior similar to that of the PMC. When the meta-crystal is covered by the ground plane, a drilled copper plate functions as an artificial electrical conductor (AEC), serving as an alternative to PEC.

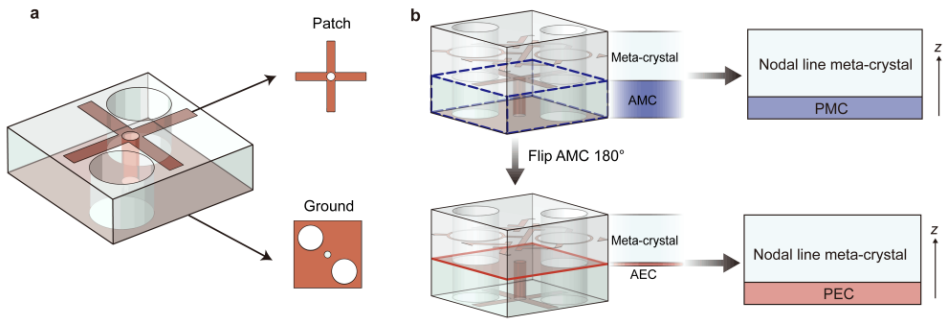

**Figure 4. Experimental boundary setup. (a)** Structure of the AMC includes metal patch and ground. **(b)** The patch side acts as the PMC and the ground side acts as the PEC.

The waveguide, composed of 5 layers of 40 × 40 cells, is sandwiched between AMC or AEC boundaries, as shown in Figures 5a and 5b (see Supporting Information Note 5 for sample details). Two chiral bulk states are observed in the measured band structures shown in Figures 5c and 5d, which align with the simulation results. By comparing the equi-frequency surfaces (EFSs) at different frequencies, we find that only one of the TM or LM appears as a chiral bulk state, depending on the boundary conditions, while the other is suppressed.

Notably, the chiral bulk states induced by the PMC-PMC boundary condition with negative dispersion show promising potential for applications such as perfect lenses, highly directional antennas, and electromagnetic invisibility. Conventional slab waveguides typically support positively dispersive modes, as described in Supporting Information Note 6. In contrast, the pseudo-gap contains only a chiral bulk state with negative dispersion, making it particularly appealing for device development.

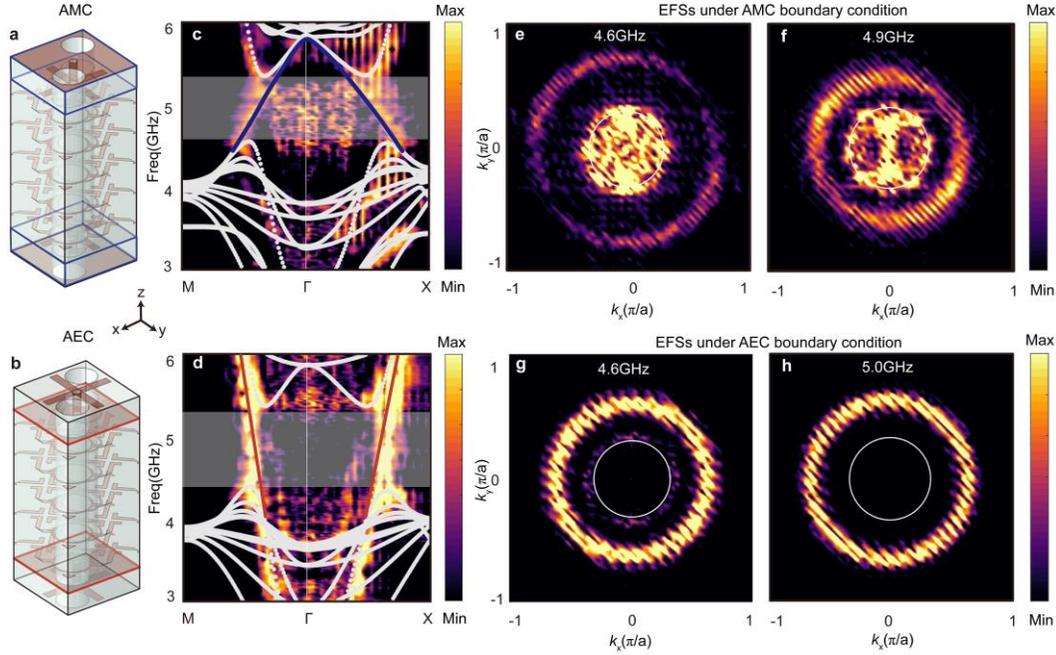

**Figure 5. Experimental observation of the chiral bulk states. (a, b)** Structure of the supercell with the **(a)** AMC and **(b)** AEC boundaries. **(c, d)** Experimentally measured bands compared with the simulated bands, with **(c)** red and **(d)** blue lines indicating the chiral bulk bands. **(e, f)** The measured EFSs under the AMC boundary condition at 4.6 and 4.9 GHz. **(g, h)** The measured EFSs under the AEC boundary condition at 4.6 and 5.0 GHz.

While considering the AEC-AMC boundary condition as shown in Figure 6a, it is evident from the simulated (Figure 6b) and measured (Figure 6c) band structures that a complete gap exists, consistent with the EFS observed at 4.75 GHz in Figure 6d. The complete gap is slightly smaller than the pseudo-gap (shadow region) under the PEC or PMC boundary condition as shown in Figures 5c and d, indicating that it arises not only from the finite size effect but also from the boundary effect. This finding provides a novel approach to manipulating a photonic gap. As shown in Supporting Information Note 7, the gap becomes shrinker as the number of layers increases but persists as long as the boundary condition remains unchanged. This phenomenon exemplifies how boundary conditions can strongly control topological chiral bulk transport properties.

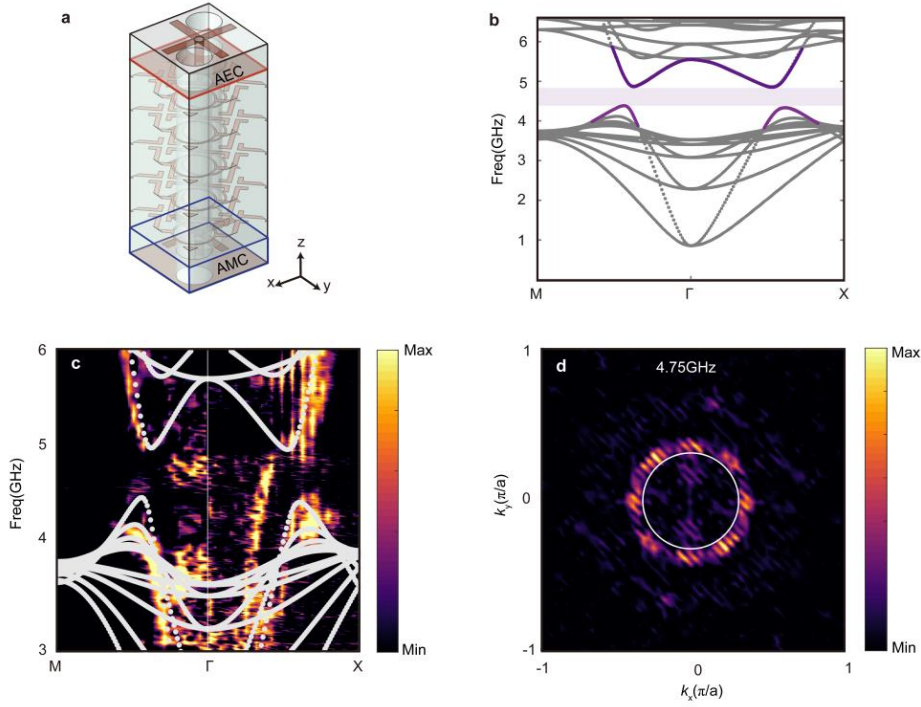

**Figure 6. Experimental measurement of the boundaries-induced gap. (a)** Supercell under the AEC-AMC boundary condition. **(b)** Simulated band structure under the PEC-PMC boundary condition. Purple bands are the hybridization of TM and LM. **(c)** Experimentally measured bands and the simulated bands (white dots) of **(a)** the supercell. **(d)** Measured EFS at 4.75 GHz, with the white circle indicating the light cone.

## Discussion

In this study, we propose that boundaries can manipulate the topological properties of bulk states and predict them analytically. In the analytical solution, we find that both the topological chiral bulk states are two chiral zeroth modes corresponding to independent zeroth-order solutions, fundamentally distinct from trivial higher-order bulk modes. We experimentally demonstrate the emergence of chiral bulk states and gaps induced by boundary effects in three-dimensional photonic nodal line meta-crystals. The PEC-PEC or PMC-PMC boundary conditions preserve mirror symmetry, and their distinct electromagnetic responses determine the chirality of bulk chiral states. Notably, our results reveal that different boundary conditions can lead to distinct chiral bulk states, irrespective of the waveguide thickness.

Essentially, the chiral bulk states originate from a nodal ring, protected by both topology and mirror symmetry. In the 3D momentum space, the nodal ring located on the $k_z = 0$ plane can be regarded as a closed loop traced by 2D Dirac points around the center. Both nodal lines and Dirac points share similar Hamiltonians, $H(k) = k_x \sigma_x + k_z \sigma_z$, and are stabilized by topological protection.

Interestingly, in the absence of an external magnetic field and adjustment of geometric parameters, only the boundary can be utilized to achieve chiral bulk states, thereby simplifying system design and fabrication and significantly enhancing practical applicability. This provides a new conceptual framework for next-generation miniature, low-loss information transmission devices, such as large-area single-mode waveguides, and filters, etc. Furthermore, these findings establish a more universal framework that is applicable to other physical systems, such as acoustics, elastics, and electronics. This broad applicability stems from the underlying physics, namely that boundaries can manipulate Dirac points, as explained in Supporting Information Note 8. Apart from unveiling a novel mechanism for electromagnetic wave chiral transmission, it also elucidates the potential of boundaries as degrees of freedom to investigate and explore the physical properties of topological semimetals.

**Method**

Our meta-crystal sample is fabricated using standard printed circuit board technology, wherein flat metal wire elements are stacked into a 3D array. In our experiments, we employed a microwave near-field scanning system to detect the chiral bulk states and the gap. To measure them accurately, we inserted a z-polarized or y-polarized dipole antenna as an excitation source into a hole located at the center of the sample's bottom surface, while another z-polarized dipole probe was inserted from the top surface into the middle layer of the sample for field distribution measurement. The field intensity and phase of the meta-crystal are measured using a vector network analyzer (VNA), providing precise point-by-point measurements with a spatial resolution of 10 mm, facilitating detailed analysis. By performing a 2D Fourier transform on the collected field map at each frequency in real space, the band structure of the measurement mode can be obtained. Supporting Information Note 9 presents experimental results showcasing the nodal ring in the bulk.

## Supporting Information

The solution process of analytical solution, a general model of boundary effects, the details of AMC and samples, simulated band structures for nodal line waveguides of varying thicknesses and parallel plate waveguides under different boundary conditions, and the experimental observation of the nodal ring can be found in the Supporting Information with Note 1-9.

## Author contributions

B.Y. conceived the idea; Y.Q. proposed the fabrication scheme; H.W. and Y.Q. and H.W. carried out all measurements; Y.Q. carried out all simulations; B.Y. developed and carried out the theoretical analysis; and B.Y. and Q.G. supervised the whole project. Y.Q. and B.Y. wrote the paper and the Supplementary Information with input from all other authors.